\documentclass[12pt]{book}

\usepackage{epstopdf,amsfonts,amsmath,amssymb}
\usepackage{graphicx}
\DeclareGraphicsExtensions{.eps}
\usepackage{appendix}

\usepackage{stackengine}
\usepackage{scalerel}
\usepackage{xcolor}

\newcommand\encadremath[1]{\vbox{\hrule\hbox{\vrule\kern8pt
\vbox{\kern8pt \hbox{$\displaystyle #1$}\kern8pt}
\kern8pt\vrule}\hrule}}
\def\enca#1{\vbox{\hrule\hbox{
\vrule\kern8pt\vbox{\kern8pt \hbox{$\displaystyle #1$}
\kern8pt} \kern8pt\vrule}\hrule}}

\newcommand\framefig[1]{
\begin{figure}[bth]
\hrule\hbox{\vrule\kern8pt
\vbox{\kern8pt \vbox{
\begin{center}
{#1}
\end{center}
}\kern8pt}
\kern8pt\vrule}\hrule
\end{figure}
}

\newcommand\figureframex[3]{
\begin{figure}[bth]
\hrule\hbox{\vrule\kern8pt
\vbox{\kern8pt \vbox{
\begin{center}
{\mbox{\epsfxsize=#1.truecm\epsfbox{#2}}}
\end{center}
\caption{#3}
}\kern8pt}
\kern8pt\vrule}\hrule
\end{figure}
}
\newcommand\figureframey[3]{
\begin{figure}[bth]
\hrule\hbox{\vrule\kern8pt
\vbox{\kern8pt \vbox{
\begin{center}
{\mbox{\epsfysize=#1.truecm\epsfbox{#2}}}
\end{center}
\caption{#3}
}\kern8pt}
\kern8pt\vrule}\hrule
\end{figure}
}

\renewcommand{\thesection}{\arabic{section}}

\makeatletter
\@addtoreset{equation}{section}
\makeatother
\newtheorem{theorem}{Theorem}[section]

\newtheorem{remark}{Remark}[section]
\newtheorem{proposition}{Proposition}[section]
\newtheorem{lemma}{Lemma}[section]
\newtheorem{corollary}{Corollary}[section]
\newtheorem{definition}{Definition}[section]

\def\br{\begin{remark}\rm\small}
\def\er{\end{remark}}
\def\bt{\begin{theorem}}
\def\et{\end{theorem}}
\def\bd{\begin{definition}}
\def\ed{\end{definition}}
\def\bp{\begin{proposition}}
\def\ep{\end{proposition}}
\def\bl{\begin{lemma}}
\def\el{\end{lemma}}
\def\bc{\begin{corollary}}
\def\ec{\end{corollary}}
\def\beaq{\begin{eqnarray}}
\def\eeaq{\end{eqnarray}}
\newcommand{\proof}{{\noindent \bf proof:}$\quad$ }
\newcommand{\eproof}{ $\square$ }

\newcommand{\be}{\begin{equation}}
\newcommand{\ee}{\end{equation}}
\newcommand{\beq}{\begin{equation}}
\newcommand{\eeq}{\end{equation}}
\newcommand{\bea}{\begin{eqnarray}}
\newcommand{\eea}{\end{eqnarray}}

\newcommand{\Tr}{\operatorname{Tr}}

\newcommand{\diag}{\operatorname{diag}}
\newcommand{\e}{{\rm e}}
\newcommand{\ii}{{\rm i}\,}

\newcommand{\CC}{{\mathbb C}}
\newcommand{\RR}{{\mathbb R}}
\newcommand{\ZZ}{{\mathbb Z}}

\newcommand{\EE}{{\mathbb E}}

\newcommand{\DD}{{\mathcal D}}

\newcommand{\Ker}{\operatorname{Ker}}

\newcommand{\Res}{\mathop{\,\rm Res\,}}

\newcommand{\td}{\tilde}
\usepackage[pdftex]{hyperref}

\usepackage{makeidx}
\makeindex
\usepackage{tocbibind}

\hypersetup{colorlinks,urlcolor=magenta,citecolor=red,linkcolor=blue,filecolor=black}

\begin{document}

\sloppy

\pagestyle{empty}
\addtolength{\baselineskip}{0.20\baselineskip}
\begin{center}
\vspace{26pt}
{\large \bf {Solutions of loop equations are random matrices.}}
\newline
\vspace{26pt}

{\sl B.\ Eynard}${}^{12}$\hspace*{0.05cm}

\vspace{6pt}
${}^{1}$ Institut de Physique Th\'{e}orique/CEA/Saclay, UMR 3681,\\
F-91191 Gif-sur-Yvette Cedex, France.\\
${}^{2}$ CRM, Centre de recherches math\'ematiques  de Montr\'eal,\\
Universit\'e de Montr\'eal, QC, Canada.
\end{center}

\begin{center}
\textbf{Abstract:}

For a given polynomial $V(x)\in \CC[x]$, a random matrix eigenvalues measure is a measure $\prod_{1\leq i<j\leq N}(x_i-x_j)^2 \prod_{i=1}^N e^{-V(x_i)}dx_i$ on $\gamma^N$. Hermitian matrices have real eigenvalues $\gamma=\RR$, which  generalize to $\gamma$ a complex Jordan arc, or actually a linear combination of homotopy classes of Jordan arcs, chosen such that integrals are absolutely convergent. Polynomial moments of such measure satisfy a set of linear equations called "loop equations".
We prove that every solution of loop equations are necessarily polynomial moments of some random matrix measure for some choice of arcs. There is an isomorphism between the homology space of integrable arcs and the set of solutions of loop equations.
We also generalize this to a 2-matrix model and to the chain of matrices, and to cases where $V$ is not a polynomial but $V'(x)\in \CC(x)$.

\end{center}


\section{Introduction}

Let us recall a few basic facts, from Mehta's book \cite{MehtaBook} for instance.

\subsection{Hermitian random matrices}

Let $V\in \RR[x]$ a real polynomial bounded from below on $\RR$ (i.e. of even degree with positive leading coefficient).
Let $\mathcal H_N$ the set of $N\times N$ Hermitian matrices, and recall that every Hermitian matrix $M\in \mathcal H_N$ can be diagonalized by a unitary conjugation
\beq
M=U X U^\dagger 
\eeq
where $U\in U(N)$ and 
\beq
X=\diag(x_1,\dots,x_N)
\eeq
is the set of its eigenvalues.
To make the decomposition unique, notice that $U$ can be right multiplied by any diagonal unitary matrix, and thus we shall consider $U$ in the quotient group $U(N)/U(1)^N$, and eigenvalues can be permuted by multiplying $U$ with a permutation matrix, eventually we roughly have
\beq
\mathcal H_N \sim (U(N)/U(1)^N \ \times \RR^N)/\mathfrak S_N.
\eeq
(remark: we abusively oversimplified the discussion, in fact when some eigenvalues are not distinct, the non-uniqueness group = the stabilizer is larger, and we should quotient $U(N)$ by the stabilizer of $X$ rather than $U(1)^N\times\mathfrak S_N$. This can be written as an orbifold, however, degenerate spectra will be of measure 0 in what follows and can be ignored).

It is well known \cite{MehtaBook} that the Lebesgue measure $\DD M$ on $\mathcal H_N$ can be rewritten as a measure on $U(N)/U(1)^N \ \times \RR^N$ as
\beq
\DD M=\prod_{i,j} dM_{i,j} =  \Delta(X)^2 \ \DD U \ \DD X
\eeq
where $\DD X=\prod_{i=1}^N dX_i$ is the Lebesgue measure on $\RR^N$ and $\DD U$ is the Haar measure on the Lie group $U(N)/U(1)^N$, and
\beq
\Delta(X) = \prod_{i<j} (x_i-x_j)
\eeq
is called the Vandermonde determinant.

A Boltzmann weight probability measure on $\mathcal H_N$ of the form
\beq
\frac{1}{\hat Z} e^{-\Tr V(M)} \DD M
\eeq
yields a marginal probability measure for eigenvalues
\beq
\frac{1}{Z} \  \Delta(X)^2 \e^{-\Tr V(X)} \DD X
\eeq
where $Z$ and $\hat Z$ are normalization factors, however, we shall from now on not normalize the measures.

Loop equations are a set of relationships (proved by integration by parts) among expectation values of symmetric polynomials of the eigenvalues, for example:
\bea\label{loopeqex1}
\EE(\Tr V'(X))=0 \cr
\forall \ k\geq 1 \qquad 
\EE(\Tr X^k V'(X)) = \sum_{j=0}^{k-1} \EE (\Tr X^j\Tr X^{k-j-1}),
\eea
and many other such relations between expectation values of product of traces of powers, that we shall detail further below.

\subsection{Generalization to normal matrices}

Let $\gamma:\RR\to \CC$ a piecewise $C^1$ Jordan arc in the complex plane.
We generalize Hermitian matrices to normal matrices (= diagonalizable by a unitary conjugation) with eigenvalues on $\gamma$:
\beq
\mathcal H_N(\gamma) = \{ M=U X U^\dagger \ | \ U\in U(N), \ X=\diag(x_1,\dots,x_N), \ x_i\in \gamma\}.
\eeq
We equip it with measure:
\beq
\DD M = \Delta(X)^2 \DD U \DD X
\eeq
where $\DD U$ is the Haar measure on $U(N)/U(1)^N$ and $\DD X=\prod_{i=1}^N dx_i$ where $dx_i$ is the curvilinear measure on $\gamma$ defined as
\beq
x_i=\gamma(s_i)  \quad \to \quad dx_i=\gamma'(s_i)ds_i
\qquad , \ s_i\in \RR
\eeq
which is in fact independent of the chosen parametrization of the Jordan arc.

For examples:

\begin{itemize}
\item $\gamma=\RR$ gives $\mathcal H_N(\RR)=\mathcal H_N$ and $\DD M$ is  the usual Lebesgue measure on $\mathcal H_N$.

\item $\gamma=S^1$ the unit circle, gives $\mathcal H_N(S^1)=U(N)$ and $\DD M$ is related to the Haar measure on $U(N)$ as
\beq
\DD M = \ii^{N^2} \ \det M^N \ \DD_{\text{Haar}(U(N))}M .
\eeq
(indeed $\ii^{-N^2}\DD M \det M^{-N}$ is a real measure, right invariant).
This formalism of normal matrices unifies Hermitian ensembles with circular ensembles (as well as many others). See  \cite{eynRMT} for examples and applications.

\end{itemize}

A Boltzmann weight measure (possibly complex) $e^{-\Tr V(M)}\DD M$ on $\mathcal H_N(\gamma)$ yields a marginal measure for the eigenvalues on $\gamma^N$:
\beq\label{lawonHNgamma}
\Delta(X)^2 e^{-\Tr V(X)} \DD X.
\eeq
Integrals of symmetric polynomials of the eigenvalues will satisfy the same loop equations \eqref{loopeqex1} as in the Hermitian case.

Notice that the measure \eqref{lawonHNgamma} can be integrated on $\gamma^N$ only for some choices of $\gamma$, namely we need the integral be absolutely convergent and thus if $\gamma$ goes to $\infty$, then  $|e^{-V(x)}|$ must tend to zero.
In order to define integrals of all symmetric polynomials of eigenvalues we shall require that $|x^k e^{-V(x)}|\to 0$ at $\infty$ on $\gamma$, for all $k\in \ZZ_+$.

In order to have the same loop equations as for the Hermitian case, we need to do integration by parts, and we need that there is no boundary term, therefore we shall require that $\gamma$ has no boundary except at $\infty$ (the case where $\gamma$ has finite boundaries at which $e^{-V(x)}\neq 0$ is called "hard edges", loop equations for hard edges can be found in \cite{eynhardedges}).

Let us now study the set of acceptable Jordan arcs for a given polynomial potential $V(x)$.
We shall study  in section  \ref{secrational} the generalization to $V'(x)\in \CC(x)$, i.e. rational case.

\section{Loop equations and measures}

\subsection{Arcs and homology}

Let $V\in \CC[x]$ a polynomial of degree $\geq 2$ written
\beq
V(x) = \sum_{k=1}^{d+1} \frac{t_k}{k}x^{k} \qquad , \ t_{d+1}\neq 0.
\eeq

Consider the set of Jordan arcs $\gamma:\RR\to\CC$, piecewise $C^1$, such that
\bea
 \gamma(-\infty)=\infty\cr
 \gamma(+\infty)=\infty\cr
 \forall k\in \ZZ_+, \   \ |x^{k} e^{-V(x)}| \ \text{bounded on } \gamma \cr
\eea
Consider the group of homotopy classes of those Jordan arcs, with addition by concatenation, and the homology space of $K$-linear combinations with $K$ a ring or field, typically $K=\ZZ$, $\mathbb Q$, $\RR$ or $\CC$.

We define
\bd
the homology space of admissible integration classes for the measure 
\bea
H_1(e^{-V(x)}dx,K)
&= & \Big\{K-\text{linear combinations of Jordan arcs }  \gamma ,\cr
&& \text{ going from }\infty \text{ to }\infty,\cr
&& \ \text{and } \ \forall \ k\geq 0,\   |x^{k} e^{-V(x)}| \text{ bounded on } \gamma \Big\} . 
\eea
It is a vector space if $K$ is a field (or a module if $K$ is a ring, let us focus on fields from now on).
\ed
The notion of integral of a holomorphic 1-form $\omega$ is well defined on a homology class $\gamma\in H_1(e^{-V(x)}dx,K) $. 
Indeed since the form is holomorphic, the integral is invariant under homotopic deformations, and for a linear combination of homotopy classes $\gamma=\sum_i c_i \gamma_i$, we define by linearity
\beq
\int_\gamma \omega \overset{\text{def}}{=} \sum_i c_i \int_{\gamma_i}\omega.
\eeq

It is clear that if $\gamma\in H_1(e^{-V(x)}dx,K)$, the integral
\beq
\int_\gamma e^{-V(x)}dx
\eeq
is absolutely convergent.

\begin{figure}
\begin{center}
\includegraphics[width=0.5\textwidth]{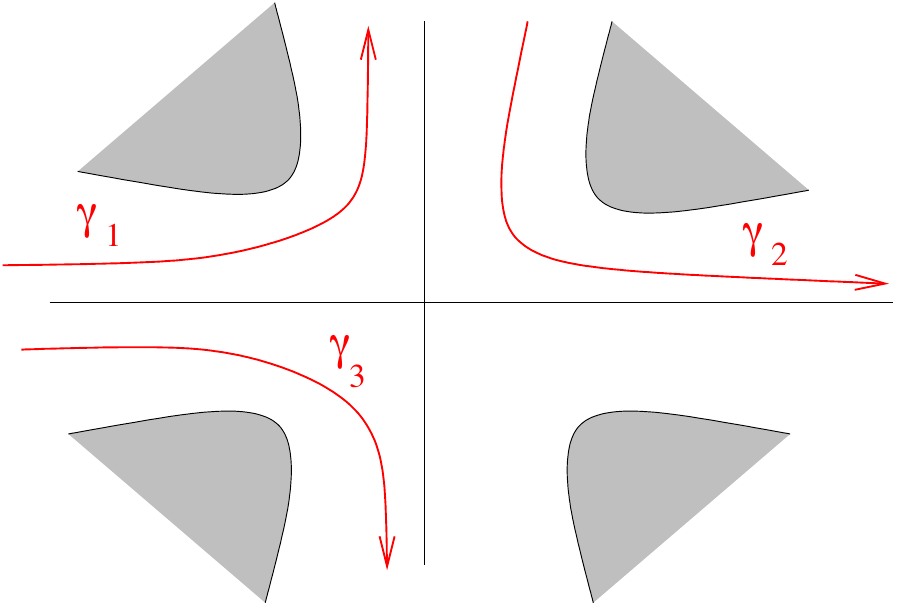}
\caption{\label{Fig1} Example with  $\deg V=4$. There are 4 sectors $\Re V>0$ and 4 sectors $\Re V<0$ near $\infty$ (shaded). Paths in $H_1$ can go from positive sector to positive sector, they must not go to $\infty$ in shaded sectors. 3 consecutive make a basis of $H_1$. For example $\RR=\gamma_1+\gamma_2$ and $\ii\RR=\gamma_1-\gamma_3$ are both in $H_1$.}
\end{center}
\end{figure}

\bp
If $\deg V=d+1\geq 2$, $H_1(e^{-V(x)}dx,K)$ has dimension
\beq
\dim H_1(e^{-V(x)}dx,K) = d.
\eeq
\ep
\proof
See \cite{Marcopath,Eynchainloop}.
Any Jordan arc going from $\infty$ to $\infty$ such that $|x^k e^{-V(x)}|$ is bounded  must start and end in sectors near $\infty$, in which $\Re V(x)\to +\infty $.
There are $d+1$ angular sectors near $\infty$ in which $\Re V(x)>0$ separated by $d+1$ sectors where $\Im V(x)<0$.
A generating family of arcs is constructed by arcs going from a sector to the next, there are $d+1$ such, and only $d$ are independent.
This is illustrated on fig.\ref{Fig1}.
\eproof

\subsection{Eigenvalues measure}

We now consider the $N$ dimensional generalization, the homology space of admissible $N$ dimensional integration domains $\subset \CC^N$, on which an $N$--dimensional spectral--matrix--model--measure is absolutely integrable, it is the symmetric $N$ tensor product:
\beq
H_N\left(\Delta(X)^2 \prod_{i=1}^N e^{-V(x_i)}dx_i,K\right)
=\text{Sym}\left(H_1\left(e^{-V(x)}dx,K\right)^{\otimes N}\right).
\eeq
Let $\gamma_1,\dots,\gamma_d$ an arbitrary basis of $H_1(e^{-V(x)}dx,K)$.
For every $d$-uple $n=(n_1,n_2,\dots,n_d)$ of non-negative integers $n_i\in\ZZ_+$ such that $\sum_{i=1}^d n_i=N$, we define
\beq
\gamma^n=
\text{sym}(\gamma_1^{n_1}\times \gamma_2^{n_2}\times \dots \times \gamma_d^{n_d})
=\frac{1}{n!} \sum_{\sigma\in \mathfrak S_N}\sigma_* \gamma_1^{n_1}\times \gamma_2^{n_2}\times \dots \times \gamma_d^{n_d}
\eeq
We may thus write
\beq
H_N\left(\Delta(X)^2 \prod_{i=1}^N e^{-V(x_i)}dx_i,K\right)
= \{ \sum_{n_1+\dots+n_d=N} c_{n_1,\dots,n_d} \text{sym}(\gamma_1^{n_1}\times \gamma_2^{n_2}\times \dots \times \gamma_d^{n_d})\}
\eeq
For short we shall call it $H_N$.

It is clear that if $\Gamma\in H_N$, the following integral
\beq
Z(\Gamma) = \int_\Gamma \Delta(X)^2 \prod_{i=1}^N e^{-V(x_i)}dx_i
\eeq
is absolutely convergent, as well as all its polynomial moments.

\bp
\beq
\dim H_N = \begin{pmatrix}
N+d-1 \cr N
\end{pmatrix} = \frac{(N+d-1)!}{N! (d-1)!}.
\eeq
\ep
\proof
This dimension is the number of $d$-uples $n=(n_1,\dots,n_d)$ such that $n_i\geq 0$ and $\sum_{i=1}^d n_i=N$.
\eproof

\subsection{Polynomial moments}

The integral $Z(\Gamma)$ is called a matrix integral, it is in fact the integral of the marginal eigenvalue distribution induced by the measure $e^{-\Tr V(M)}\DD M$ on $\mathcal H_N(\Gamma)$.

Let $\mathcal P_N=\CC[x_1,\dots,x_N]^{\text{Sym}}$ the vector space of all symmetric polynomials of $N$ variables.

\bd
For $\Gamma\in H_N$, the measure $\Delta(X)^2 \prod_i e^{-V(x_i)}dx_i$ defines the following map:
\bea
\EE_{\Gamma}:  & \mathcal P_N & \to \CC \cr
& p & \mapsto \int_\Gamma p(x_1,\dots,x_N) \ \Delta(X)^2 \prod_{i=1}^N e^{-V(x_i)}dx_i
\eea
which is a linear form on $\mathcal P_N$:
\beq
\EE_\Gamma\in \mathcal P_N^*.
\eeq
\ed
 
Since $H_N$ is a vector space, and the map $\EE:\Gamma\mapsto \EE_\Gamma$ is clearly linear, we have a homeomorphism of vector spaces.
A key result is that this homeomorphism is injective:
 
\bt[Injectivity]\label{thinjective}
$\EE$ is an injective homeomorphism of vector spaces
\bea
\EE: & H_N &\to \mathcal P_N^* \cr
& \Gamma & \mapsto \EE_\Gamma
\eea 
 
\et

\proof
The full algebraic proof is given in Appendix~\ref{Appproofinjective}.
We sketch its argument here. Let $\Gamma\in\Ker\EE$ and write
\beq
\Gamma= \sum_{n=(n_1,\dots,n_d), \ n_1+\dots+n_d=N} c_{n} \gamma^n,
\eeq
Let $I_\Gamma$ be the ideal of polynomials $f$ for which
\beq
\int_\Gamma f(X)q(X)e^{-\Tr V(X)}\,dX=0
\qquad\forall q\in\CC[x_1,\ldots,x_N].
\eeq
Symmetrizing the test polynomials shows that $\Delta(X)^2\in I_\Gamma$.
Since $V'$ is polynomial, integration by parts gives
\beq
\partial_i I_\Gamma\subseteq I_\Gamma,
\qquad 1\le i\le N.
\eeq
Repeated differentiation of $\Delta(X)^2$ therefore gives $1\in I_\Gamma$.
Thus all polynomial moments of $\Gamma$ for the product weight
$e^{-\Tr V(X)}\,dX$ vanish. One-variable injectivity and its tensor
power then imply that every $c_n$ vanishes, hence $\Gamma=0$.
\eproof

\subsubsection{Symmetric polynomials}

Let the power sums be defined as the following symmetric polynomials:
\beq
p_k(x_1,\dots,x_N) = \sum_{i=1}^N x_i^k = \Tr X^k.
\eeq
For $\mu=(\mu_1\geq\mu_2\geq \mu_3\geq \dots \geq \mu_\ell)$ a partition, we denote
\beq
p_\mu(x_1,\dots,x_N) = \prod_{j=1}^\ell p_{\mu_j}(x_1,\dots,x_N).
\eeq
We shall also use the same notation when $\mu=(\mu_1,\dots,\mu_\ell)$ is a $\ell$--uple (no ordering assumed).
We recall the notations:
\begin{itemize}
\item weight of a partition (resp. a upple)
\beq
|\mu| = \sum_{i} \mu_i,
\eeq
\item length of a partition (resp. a upple)
\beq
\ell(\mu) = \# \{ i \ | \ \mu_i\neq 0\}.
\eeq
\end{itemize}

We recall the classical lemma:
\bl[Basis of $\mathcal P_N$]\label{lemmaPN}
A basis of $\mathcal P_N$ is given by 
\beq
\{ p_\mu \ | \ \ell(\mu)\leq N\}.
\eeq
\el
\proof{Easy by recursion on $N$. See appendix \ref{applemmaPN}.}

\subsection{Loop equations}

Define
\bd
For a $n$-uple $\mu=(\mu_1,\dots,\mu_n)$ (not necessarily ordered), let the following symmetric polynomial
\bea
Q_\mu 
&=& \sum_{j=0}^{d} t_{j+1} p_{\mu_1+j}p_{\mu_2}\dots p_{\mu_n} \cr
&& - \sum_{j=0}^{\mu_1-1} p_j p_{\mu_1-1-j}p_{\mu_2}\dots p_{\mu_n} \cr
&& - \sum_{i=2}^n \mu_i \ p_{\mu_1+\mu_i-1} \prod_{\substack{2\le k\le n\\k\neq i}} p_{\mu_k}
\eea
They generate
\beq
\mathcal L
= \text{Span}\left< Q_\mu\right>_{n, \ \mu=(\mu_1,\dots,\mu_n)} \subset \mathcal P_N.
\eeq
\ed

\bd[Loop equations]
A linear form
$E\in \mathcal P_N^*$ is called a solution of loop equations iff
\beq
E(\mathcal L)=0.
\eeq
The set of solutions of loop equations is denoted
\beq
\mathcal L^\perp = \{E\in \mathcal P_N^* \ | \ \forall p\in \mathcal L, \ E(p)=0\}.
\eeq
\ed

\bt[Matrix integrals satisfy loop equations]
\beq
\forall \Gamma\in H_N \ \qquad , \quad \EE_\Gamma\in \mathcal L^\perp.
\eeq
The map $\EE:H_N\to \mathcal L^\perp$ is an injective homeomorphism, and
\beq
\dim \mathcal L^\perp \geq \frac{(N+d-1)!}{N! (d-1)!}.
\eeq
\et

\proof
This is a well known theorem in random matrix theory, it is a special case of Schwinger-Dyson equations (Schwinger-Dyson equations are more generally defined for quantum field theories (QFT)). When the QFT is a matrix integral, these were called "loop equations" by Migdal  \cite{Migdalloop}. Schwinger-Dyson equations merely reflect the fact that an integral is invariant under change of variable. They can also be rewritten as just integration by parts \cite{David,eynRMT}.
Indeed notice that
\beq
\begin{aligned}
&Q_\mu(X)\Delta(X)^2e^{-\Tr V(X)}\\
&\qquad=-\sum_{i=1}^N\frac{\partial}{\partial x_i}
\left(x_i^{\mu_1}\prod_{k=2}^n p_{\mu_k}(X)\,
\Delta(X)^2e^{-\Tr V(X)}\right),
\end{aligned}
\eeq
immediatly implying that
\beq
\EE_\Gamma(Q_\mu)=0.
\eeq
This relies on the fact that the integrand vanishes at the boundaries of $\Gamma$, i.e. at $\infty$.
See \cite{eynhardedges} in case there would be boundary terms.
\eproof

\subsection{Every solution of loop equations is a matrix integral}

The morphism $\EE$ is in fact an isomorphism.
This means that to every solution $E$ of loop equations corresponds a $\Gamma\in H_N$ such that $E=\EE_\Gamma$.
The following is the main theorem of this article

\bt[Solutions of loop equations = matrix eigenvalues integrals]\label{mainth1}
The map $\EE:H_N\to \mathcal L^\perp$ is an isomorphism:
\beq
\dim \mathcal L^\perp = \dim H_N = \frac{(N+d-1)!}{N! (d-1)!}.
\eeq

\et

\proof
We need to prove surjectivity.
Let $E\in \mathcal L^\perp$.

\begin{figure}
\begin{center}
\includegraphics[width=0.2\textwidth]{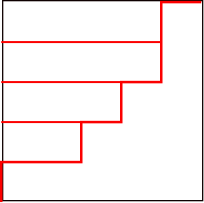}
\caption{\label{Fig2} An element of $A_{N,d}$ can be seen as a Ferrer diagram that can fit in a box $(d-1)\times N$. It is thus a line of length $N+d-1$ with $N$ vertical steps and $d-1$ horizontal steps. The number of such lines is the number of ways to choose $N$ vertical steps among $N+d-1$.}
\end{center}
\end{figure}

Let
\beq
A_{N,d} = \{
\mu=\text{partitions}\ , \ \ell(\mu)\leq N \ \text{and} \ \forall i , \ \mu_i\leq d-1
\}.
\eeq
We have (see fig.\ref{Fig2})
\beq
\# A_{N,d} = \frac{(N+d-1)!}{N! (d-1)!}.
\eeq

We shall first prove the following lemma:
\bl
There exist some coefficients $c_{\mu,\nu}$ such that for every partition $\mu$
\beq
E(p_\mu) = \sum_{\nu\in A_{N,d}} c_{\mu,\nu} E(p_\nu).
\eeq
\el
\textbf{Proof of the lemma:}
We prove it by recursion on $k=|\mu|$.
It clearly holds for $k=0$ since the empty partition is already in $A_{N,d}$.
Assume that it holds up to $k-1$.
Let $\mu$ a partition of weight $|\mu|=k$ and length $\ell(\mu)\leq N$.
If all $\mu_i\leq d-1$ then $\mu\in A_{N,d}$ so the lemma holds.
If there exists $\mu_i\geq d$, up to relabeling assume that it is $\mu_1\geq d$.
The loop equation $E(Q_{\mu_1-d,\mu_2,\dots\mu_n})=0$ implies
\bea
t_{d+1} E(p_\mu) 
&=& -\sum_{j=0}^{d-1} t_{j+1} E(p_{\mu_1-d+j}p_{\mu_2}\dots p_{\mu_n}) \cr
&& + \sum_{j=0}^{\mu_1-d-1} E(p_j p_{\mu_1-d-1-j}p_{\mu_2}\dots p_{\mu_n}) \cr
&& + \sum_{i=2}^n \mu_i \ E(p_{\mu_1-d+\mu_i-1} \prod_{\substack{2\le k\le n\\k\neq i}} p_{\mu_k}).
\eea
We recall that $t_{d+1}\neq 0$
and we notice that all polynomials in the right hand side have weights $<k$.
By the recursion hypothesis, this implies that all terms in the right hand side are linear combinations of $E(p_\nu)$ with $\nu\in A_{N,d}$.
If $\mu$ is such that $\ell(\mu)>N$, according to lemma \ref{lemmaPN}, we can rewrite $p_\mu$ as a linear combination of $p_\nu$s of the same weight $|\nu|=|\mu|$ with $\ell(\nu)\leq N$.
This ends the proof of the lemma.\eproof

This implies that the map $E:\mathcal P_N\to \CC$ is entirely determined by its value on the subspace
\beq
\text{span}\left< p_\mu\right>_{\mu\in A_{N,d}}
\eeq 
and therefore
\beq
\dim \mathcal L^\perp \leq \# A_{N,d}  = \frac{(N+d-1)!}{N! (d-1)!}.
\eeq
Since we already knew the opposite inequality this implies equality:
\beq
\dim \mathcal L^\perp  = \frac{(N+d-1)!}{N! (d-1)!},
\eeq
which thus implies that $\EE$ is an ismorphism.
\eproof

\section{2-Matrix model}

\subsection{Setting, arcs and homology}
Consider 2 random normal matrices $M,\td M$ of size $N\times N$, with eigenvalues on some arcs $\gamma$ and $\td\gamma$, i.e. in $\mathcal H_N(\gamma)\times \mathcal H_N(\td\gamma)$, with a measure
\beq
e^{-\Tr (V(M)+\td V(\td M) - M\td M)} \DD M \ \DD\td M,
\eeq
where $V$ and $\td V$ are polynomials of respective degrees $d+1$ and $\td d+1$, written
\beq
V(x) = \sum_{k=1}^{d+1} \frac{t_k}{k}x^k
\quad , \quad
\td V(y) = \sum_{k=1}^{\td d+1} \frac{\td t_k}{k}y^k.
\eeq

On a product contour $\gamma^N\times\td\gamma^N$, diagonalizing
$M=U X U^\dagger$ and $\td M=\td U Y \td U^\dagger$ and using
the Harish-Chandra--Itzykson--Zuber integral over $U(N)$
\cite{MehtaBook,eynRMT} gives, up to a constant depending only on $N$,
the marginal measure
\beq
\Delta(X)\Delta(Y) \det(e^{x_iy_j}) \ \prod_{i=1}^N e^{-V(x_i)}dx_i \prod_{i=1}^N e^{-\td V(y_i)}dy_i.
\eeq
For general contours, we retain the pairing between $x_i$ and $y_i$
and use instead the measure
\beq
\Delta(X)\Delta(Y)
\prod_{i=1}^N e^{x_i y_i-V(x_i)-\td V(y_i)}\,dx_i\,dy_i.
\eeq
It is invariant under simultaneous permutations of the pairs
$(x_i,y_i)$. Integration contours are symmetrized under these
simultaneous permutations. The contours must make all polynomial
moments absolutely convergent and all integration-by-parts boundary
terms vanish.

The two expressions agree on product contours, up to the factor $N!$:
for every $p\in\mathcal P_N$,
\beq
\begin{aligned}
&\int_{\gamma^N\times\td\gamma^N}
p(X)\Delta(X)\Delta(Y)\det(e^{x_i y_j})
e^{-\Tr V(X)-\Tr\td V(Y)}\,dX\,dY\\
&\qquad=N!\int_{(\gamma\times\td\gamma)^N}
p(X)\Delta(X)\Delta(Y)
\prod_{i=1}^N e^{x_i y_i-V(x_i)-\td V(y_i)}\,dx_i\,dy_i.
\end{aligned}
\eeq
Indeed, after expanding the determinant, a permutation of the
$y$-variables makes every term equal; its sign cancels the sign
of $\Delta(Y)$. This argument uses the product contour and does
not replace the paired exponential by a determinant on an arbitrary
paired contour.

\bl
If $d\td d> 1$, 
we have
\beq
H_1(e^{xy-V(x)-\td V(y)}dxdy) = H_1(e^{-V(x)}dx)\otimes H_1(e^{-\td V(y)}dy)
\eeq
thus
\beq
\dim H_1 =d\td d,
\eeq
and a basis of $H_1$ is made of products $\gamma_{i,j}:=\gamma_i\times\td\gamma_j$.
For the paired measure we define
\beq
H_N=\operatorname{Sym}^N H_1
=\operatorname{Sym}^N\!\left(
H_1(e^{-V(x)}dx)\otimes H_1(e^{-\td V(y)}dy)\right).
\eeq
Its dimension is
\beq
\dim H_N
= \frac{(N+d\td d-1)!}{N! (d\td d-1)!}
\eeq
A basis is given by
\beq
\left\{\prod_{i,j}\gamma_{i,j}^{n_{i,j}}:
n_{i,j}\ge0,\ \sum_{i,j}n_{i,j}=N\right\},
\eeq
where each product denotes symmetrization of the pairs.
\el
\proof
For $d\td d>1$, admissible product contours give the one-pair
homology and its dimension; see \cite{Bertola_2003,Marcopath}.
Taking its symmetric tensor power gives the displayed occupation
basis and dimension.
\eproof

\subsection{Moments linear map }

Let $\Gamma\in H_N$, we define the linear map
\bea
\EE_\Gamma \ : & \mathcal P_N & \to \CC \cr
& p & \mapsto \displaystyle\int_\Gamma p(X)\Delta(X)\Delta(Y)
\prod_{i=1}^N e^{x_i y_i-V(x_i)-\td V(y_i)}\,dx_i\,dy_i.
\eea
The integration defines a morphism
\bea
\EE \ : & H_N & \to  \mathcal P_N^*  \cr
& \Gamma & \mapsto \EE_\Gamma
\eea

In what follows we shall need to extend $E_\Gamma$ to define the mixed matrix insertions.
Following \cite{EynPrats}, for an invariant polynomials $F(M,\td M)$ (satisfying jointly $F(UMU^\dagger,U\td M U^\dagger)=F(M,\td M)$  $\forall \ U\in U(N)$), we define $\EE_\Gamma(F)$ by a Gaussian integral over
strictly upper triangular complex matrices $T$:
\beq\label{eqpairedGaussianlift}
\begin{aligned}
\EE_\Gamma(F)
={}&\pi^{-N(N-1)/2}\int_\Gamma\Delta(X)\Delta(Y)
\prod_{i=1}^N e^{x_i y_i-V(x_i)-\td V(y_i)}\\
&\quad\times\left(\int F(X+T,Y-T^\dagger)
e^{-\Tr TT^\dagger}\,dT\right)dX\,dY.
\end{aligned}
\eeq
Here $dT=\prod_{a<b}d\Re T_{a,b}\,d\Im T_{a,b}$, so the normalized
Gaussian integral of $1$ is $1$. Since the traces of powers of an
upper triangular matrix depend only on its diagonal, this extension
agrees with the preceding definition for $p\in\mathcal P_N$.

This is the Schur representation of a matrix integration cycle,
\[
(M,\td M)=
\bigl(U(X+T)U^\dagger,\ U(Y-T^\dagger)U^\dagger\bigr),
\qquad U\in U(N).
\]
Modulo the diagonal unitary redundancy, the pullback of the
holomorphic matrix volume form is a nonzero
constant times $\Delta(X)\Delta(Y)\,dX\,dY\,dT\,\DD U$.
Moreover,
\[
\Tr\bigl(V(M)+\td V(\td M)-M\td M\bigr)
=\Tr V(X)+\Tr\td V(Y)-\Tr XY+\Tr TT^\dagger.
\]
See \cite{EynPrats} for the Schur and triangular-integral formulas.
Polynomial changes of the matrices therefore give the usual
matrix integration-by-parts identities. Explicitly, for a matrix
word $F$ multiplied by any product of traces,
\beq\label{eqpairedWard}
\begin{aligned}
\EE_\Gamma\!\left(\Tr\bigl((V'(M)-\td M)F\bigr)\right)
&=\EE_\Gamma\!\left(\sum_{a,b}
\frac{\partial F_{a,b}}{\partial M_{a,b}}\right),\\
\EE_\Gamma\!\left(\Tr\bigl((\td V'(\td M)-M)F\bigr)\right)
&=\EE_\Gamma\!\left(\sum_{a,b}
\frac{\partial F_{a,b}}{\partial\td M_{a,b}}\right).
\end{aligned}
\eeq
The Gaussian factor and the admissible contours remove all boundary
terms; the angular variables have no boundary. These statements
are linear in $\Gamma$ and thus hold on all of $H_N$.

\bt[Injectivity]
The morphism $\EE: H_N \to \mathcal P_N^*  $ is injective.
\et

\proof
Suppose that $\EE_\Gamma(p)=0$ for every $p\in\mathcal P_N$.
First, all products of mixed traces also have zero expectation.
Indeed, induct on the total number of occurrences of $\td M$.
The initial case is our assumption. In a product containing $\td M$,
cyclically write one trace as $\Tr(\td M F)$ and apply the first
identity in \eqref{eqpairedWard}, including the other traces in $F$.
Both $\Tr(V'(M)F)$ and the matrix divergence have fewer occurrences
of $\td M$, so the induction applies.

Next, these mixed insertions determine all polynomial moments of
the paired eigenvalues. The polynomials
$\sum_i x_i^a y_i^b$ generate the polynomials invariant under
simultaneous permutations of the pairs. For a product of such sums,
take the corresponding product of traces $\Tr(M^a\td M^b)$.
Its normalized Gaussian integral in \eqref{eqpairedGaussianlift}
has this product as its highest total degree term. Every other
term contains a Gaussian contraction of two triangular entries
and has lower total degree. After simultaneous symmetrization,
induction on total degree therefore gives
\beq\label{eq:eqqxy2mm0}
\int_\Gamma q(X,Y)\Delta(X)\Delta(Y)
\prod_i e^{x_i y_i-V(x_i)-\td V(y_i)}\,dX\,dY=0
\eeq
for every polynomial $q$ invariant under simultaneous permutations
of the pairs. 
Since both the cycle $\Gamma$ and the weight are
invariant under these permutations, the integral of an arbitrary
polynomial $q$ equals that of its symmetrization, therefore \eqref{eq:eqqxy2mm0} holds  for every polynomial $q$.

Then apply the ideal argument of Appendix~\ref{AppproofinjectivePol}
to the product weight $\prod_i e^{x_i y_i-V(x_i)-\td V(y_i)}$,
with $\Delta(X)\Delta(Y)$ in place of $\Delta(X)^2$.
The phase is polynomial, so its annihilator ideal is stable under
all $\partial_{x_i}$ and $\partial_{y_i}$. It contains
$\Delta(X)\Delta(Y)$, and repeated differentiation gives $1$.
Thus all moments of $\Gamma$ for the product weight vanish.
The one-pair polynomial period pairing is nondegenerate by
\cite[Theorem~3.1]{Bertola_2003}. Taking tensor powers therefore
implies $\Gamma=0$.
\eproof

\subsection{Loop equations}

The loop equations of the 2-matrix model are slightly more subtle, they can be found for example in \cite{Eynchainloop,eynhardedges}.

For comparison with the HCIZ representation, and to retain the
notation used in Section~4, define the $N\times N$ matrix $U(X,Y)$ by
\beq
U(X,Y)_{i,j} = e^{x_i y_j}
\eeq
then define
\beq
R_{i,j}(X,Y) = e^{x_i y_j} (U(X,Y)^{-1})_{j,i}
\quad , \quad 
R_i^{(l)}(X,Y) = \sum_j R_{i,j}(X,Y) y_j^l .
\eeq
Notice that
\beq
\sum_i R_{i,j}(X,Y)=1 \quad , \quad \sum_j R_{i,j}(X,Y)=1
\quad , \quad R^{(0)}_{i}(X,Y)=1.
\eeq

These HCIZ functions are auxiliary notation. For the paired
functional, the mixed insertions are matrix traces, evaluated by
\eqref{eqpairedGaussianlift}. We write
\beq
p^{(l)}_k=\Tr(M^k\td M^l),
\qquad p^{(0)}_k=p_k(X).
\eeq
In particular, $p^{(l)}_k$ is not defined by the preceding HCIZ
functions on an arbitrary paired contour.

\bt[Loop equations]\label{thloopeq2MM}
For each $n$--uple $(\mu_1,\mu_2,\dots,\mu_n)$, there is a symmetric polynomial $Q_{\mu_1,\mu_2,\dots,\mu_n}(X) \in \mathcal P_N$, of highest weight term
\beq
\begin{aligned}
Q_{\mu_1,\ldots,\mu_n}(X)
={}&\td t_{\td d+1}(t_{d+1})^{\td d}
p_{\mu_1+d\td d,\mu_2,\ldots,\mu_n}(X)\\
&+\sum_{\nu,\ |\nu|<|\mu|+d\td d}c_{\mu,\nu}p_\nu(X),
\end{aligned}
\eeq
such that
\beq
\EE_\Gamma(Q_{\mu_1,\mu_2,\dots,\mu_n}(X))=0,
\eeq

\et

\proof
This is the matrix integration-by-parts calculation of
\cite{Eynchainloop,eynhardedges}, applied to the paired functional.
In the first identity of \eqref{eqpairedWard}, take
$F=M^k\td M^l\prod_{i=2}^n p_{\mu_i}(X)$.
In the second, take $F=M^k\prod_{i=2}^n p_{\mu_i}(X)$.
Computing the matrix divergences gives
\beq
\begin{aligned}
&\EE_\Gamma\!\left(p^{(l+1)}_k\prod_{i=2}^n p_{\mu_i}\right)\\
&\quad=\sum_{j=0}^d t_{j+1}
\EE_\Gamma\!\left(p^{(l)}_{k+j}\prod_{i=2}^n p_{\mu_i}\right)\\
&\qquad-\sum_{j=0}^{k-1}
\EE_\Gamma\!\left(p^{(l)}_j p_{k-1-j}\prod_{i=2}^n p_{\mu_i}\right)\\
&\qquad-\sum_{i=2}^n\mu_i
\EE_\Gamma\!\left(p^{(l)}_{k+\mu_i-1}
\prod_{\substack{2\le h\le n\\h\ne i}}p_{\mu_h}\right),
\end{aligned}
\eeq
\beq
\EE_\Gamma\!\left(p_{k+1}\prod_{i=2}^n p_{\mu_i}\right)
=\sum_{l=0}^{\td d}\td t_{l+1}
\EE_\Gamma\!\left(p^{(l)}_k\prod_{i=2}^n p_{\mu_i}\right).
\eeq
The first equation is a recursion on $l$, with initial condition
\beq
\EE_\Gamma\!\left(p^{(0)}_k\prod_{i=2}^n p_{\mu_i}\right)
=\EE_\Gamma\!\left(p_k\prod_{i=2}^n p_{\mu_i}\right),
\eeq
which expresses every $\EE_\Gamma(p^{(l)}_k\prod_{i=2}^n p_{\mu_i})$
as a linear combination of first-matrix moments.
Substituting these expressions into the second identity gives
\beq
\EE_\Gamma(Q_{k,\mu_2,\dots,\mu_n}(X))=0,
\eeq
where $Q_{k,\mu_2,\dots,\mu_n}(X)$ is a symmetric polynomial of $x$, thus a linear combination of power sums symmetric polynomials.
At each step in $l$, the highest weight term comes from $t_{d+1}$;
all matrix-divergence terms have lower weight. Since $d\td d>1$,
the term $p_{k+1}$ in the second identity also has lower weight.
Thus the highest weight term is
\beq
\begin{aligned}
Q_{k,\mu_2,\ldots,\mu_n}(X)
={}&\td t_{\td d+1}(t_{d+1})^{\td d}
p_{k+d\td d,\mu_2,\ldots,\mu_n}(X)\\
&+\sum_{\nu,\ |\nu|<k+d\td d+\sum_{i=2}^n\mu_i}
c_{(k,\mu_2,\ldots,\mu_n),\nu}p_\nu(X).
\end{aligned}
\eeq

\eproof

\bd[Loop equations]
We define the loop equations sub-space
\beq
\mathcal L = \text{Span}\left<Q_{\mu_1,\mu_2,\dots,\mu_n}(X) \right> \subset \mathcal P_N,
\eeq
and the set of solutions of loop equations
\beq
\mathcal L^\perp = \{ E\in \mathcal P_N^* \ | \ E(\mathcal L)=0\} \subset \mathcal P_N^*.
\eeq
\ed

We have proved that
\bp
The map $\EE: H_N\to \mathcal L^\perp$, $\Gamma\mapsto \EE_\Gamma$, is an injective homeomorphism.
We thus have
\beq
\dim \mathcal L^\perp \geq \frac{(N+d\td d-1)!}{N! (d\td d-1)!}.
\eeq

\ep

\subsection{Solutions of loop equations are matrix integrals}

In fact the map is an isomoprhism
\bt[Solutions of loop equations = matrix integrals]\label{mainth2MM}
The map $\EE: H_N\to \mathcal L^\perp$, $\Gamma\mapsto \EE_\Gamma$, is an isomorphism.
\beq
\dim \mathcal L^\perp = \frac{(N+d\td d-1)!}{N! (d\td d-1)!}.
\eeq

\et

\proof
By Lemma~\ref{lemmaPN}, it suffices to consider partitions with
$\ell(\mu)\le N$.
We prove that $E\in \mathcal L^\perp$ is determined by its value on the subspace
\beq
\text{Span}\left< p_\mu \right>_{\mu\in A_{N,d\td d}}.
\eeq
In other words we show that for any partition $\mu$:
\beq\label{eqmuintermsofAndd2MM}
E(p_\mu) = \sum_{\nu\in A_{N,d\td d}} c_{\mu,\nu} E(p_\nu).
\eeq
This is proved by recursion on $|\mu|$. This is obviously true when $|\mu|=0$, assume it is true up to $|\mu|-1$.
If $\mu\notin A_{N,d\td d}$, this means that one row, let us say $\mu_1\geq d\td d$, the loop equation
\beq
E(Q_{\mu_1-d\td d,\mu_2,\dots,\mu_n})=0
\eeq
implies \eqref{eqmuintermsofAndd2MM}.

This implies that
\beq
\dim \mathcal L^\perp \leq \# A_{N,d\td d} = \frac{(N+d\td d-1)!}{N! (d\td d-1)!},
\eeq
and since we already have the opposite inequality from injectivity, we conclude that there is equality and $\EE$ is an isomorphism.
\eproof

\section{Chain of normal matrices}

The chain of matrices is for example defined in \cite{MehtaBook,eynmetha,Eynchainloop}.

Consider some complex polynomials  $V_1,V_2,\dots,V_L$, of respective degrees
\beq
\deg V'_i=d_i.
\eeq
Consider the measure
\beq
\DD P(X_1,\dots,X_L) = \Delta(X_1)\Delta(X_L) \prod_{l=1}^{L-1} \det_{a,b}(e^{ (X_l)_a (X_{l+1})_b}) \ \prod_{l=1}^L e^{-\Tr V_l(X_l)} \DD X_l
\eeq
that we shall put on $\Gamma\in H_N$:
Let
\beq
H_N = \otimes_{l=1}^L H_N(\Delta(X_l)^2 e^{-\Tr V(X_l)}\DD X_l) .
\eeq
We have
\beq
\dim H_N = D=\prod_{l=1}^L d_l.
\eeq

For $\Gamma\in H_N$ we define
\bea
\EE_\Gamma : \ & \mathcal P_N &\to \CC \cr
& p &\mapsto \int_\Gamma p(X_1) \DD P(X_1,\dots,X_L).
\eea
and the map
\bea
\EE : \ & H_N & \to \mathcal P_N^*  \cr
& \Gamma &\mapsto \EE_\Gamma.
\eea
This map is injective, the proof is more or less the same as the 1 matrix model.

\subsubsection{Loop equations}

Define
\bea
p^{(l_2,\dots,l_L)}_k(X_1,X_2,\dots,X_L)
&=& \sum_{i_1,\dots,i_L} ((X_1)_{i_1})^k R(X_1,X_2)_{i_1,i_2} ((X_2)_{i_2})^{l_2} \cr && R(X_2,X_3)_{i_2,i_3} \dots R(X_{L-1},X_L)_{i_{L-1},i_L} ((X_L)_{i_L})^{l_L}.
\eea

\bt[Loop equations]\label{thloopeq2MM}
For each $n$--uple $(\mu_1,\mu_2,\dots,\mu_n)$, there is a symmetric polynomial $Q_{\mu_1,\mu_2,\dots,\mu_n}(X) \in \mathcal P_N$, of highest weight term
\beq
Q_{\mu_1,\mu_2,\dots,\mu_n}(X)
= C \ p_{\mu_1+D,\mu_2,\dots,\mu_n}(X) + \sum_{\nu, \ |\nu|<|\mu|+D} c_{\mu,\nu} p_\nu(X),
\eeq
with $C\neq 0$,
such that
\beq
\EE_\Gamma(Q_{\mu_1,\mu_2,\dots,\mu_n}(X))=0,
\eeq
We denote $\mathcal L=\text{Span}<Q_\mu>$, and $\mathcal L^\perp=\{ E\in \mathcal P_N^* \ | \ E(\mathcal L)=0\}$.
\et

This theorem was proved in \cite{Eynchainloop}.
The coefficient $C$ is the leading doefficient of $V'_L\circ V'_{L-1} \circ \dots \circ V'_1$.

The rest is the same as for 1 and 2 matrix models:
\bt
$\EE$ is an isomorphism
\bea
\EE \ : \quad & H_N & \to \mathcal L^\perp \cr
& \Gamma & \mapsto \EE_\Gamma
\eea
is an isomorphism
\beq
\dim \mathcal L^\perp = \dim H_N = \frac{(N+D-1)!}{N! (D-1)!}
\eeq
where $D=\prod_{l=1}^L \deg V'_l$.
\et
The proof is exactly the same as 1 and 2 matrix models.

\section{Rational potentials and hard edges}
\label{secrational}

Now we will consider 
\beq
V'(x)\in \CC(x)
\eeq
which means that $V(x)$ can also have logarithms. This is the semi-classical   of  Bertola See \cite{Marcopath, Bertola_2003}.
The degree of $V'(x)$ is defined to be the sum of degrees of all poles, including the pole at $\infty$.
\beq
\deg V' = \sum_{p=\text{poles}} \deg_p V'.
\eeq
Notice that $e^{-V(x)}$ has essential singularities at pole of $V'(x)$, and if $V'$ has a simple pole $p$ with a non-vanishing residue $r=\Res_p V'$, 3 situations can occur:
\begin{itemize}
\item $r\in \ZZ_-$: then $e^{-V(x)}$ has a zero at $p$ or order $r$.
\item $r\in \ZZ_+$: then $e^{-V(x)}$ has a pole at $p$ of degree $r$.
\item $r\notin \ZZ$: then $e^{-V(x)}$ is not analytic at $p$, we need to introduce a cut ending at $p$.
\end{itemize}
Let us consider the complex plane from which we remove all poles, and possibly cuts ending at poles, so that $e^{-V}$ is analytic in the considered domain.

The admissible Jordan arcs, are now arcs going from a pole to another (or the same pole), and not crossing cuts. 
Arcs can arrive at a pole only in a direction in which $\Re V(x)\to +\infty$.

$\bullet$ If $e^{-V(x)}$ has a zero, an arc can end on it from any direction.

$\bullet$ If $e^{-V(x)}$ has a pole, no arc can end on it, but can go around it, for instance a small closed circle around a pole is an admissible arc.

$\bullet$ If $e^{-V(x)}$ has a cut, arcs must go around the cut without crossing it.

\smallskip
For the homology dimension formulas and the one-matrix statements below,
we assume that $V'$ has a pole at $\infty$. Equivalently, writing
$V'=R/D$ as a reduced fraction, we assume $\deg R>\deg D$.
We work over $\CC$ in the one-matrix statements. Some of the classical
examples listed below lie outside this growth assumption and require
a separate homology count at infinity.
Under this assumption, the arcs described in
\cite{Marcopath, Eynchainloop} have $\deg V'$ independent homology classes.

\subsubsection{Hard edges and semi-classical potentials}

Let us also consider $l$ distinct generic points $b_1,\dots,b_l$ (distinct from the poles of $V'$), and let us also consider some paths $\gamma$ that can end at some $b_i$.
These are called hard edges.
The presence of hard-edges will increase the dimension of $H_1$ by $l$.

\subsubsection{Semi-classical potentials}

\begin{itemize}
\item Legendre
\beq
e^{-V(x)} = 1 \qquad  \text{hard edges } = \gamma=[-1,1].
\eeq

\item Chebyshev
\beq
e^{-V(x)} = \frac{1}{\sqrt{x_2-1}}  \qquad  \gamma = [-1,1].
\eeq

\item Hermite
\beq
e^{-V(x)} = e^{-x^2}  \qquad  \gamma = \RR.
\eeq

\item Laguerre, for $\alpha>0$
\beq
e^{-V(x)} = a^{\alpha-1} e^{-x}  \qquad  \gamma = \RR_+.
\eeq

\item Jacobi, for $\alpha,\beta>0$
\beq
e^{-V(x)} = (1-x)^{\alpha-1}(1+x)^{\beta-1}  \qquad  \gamma = [-1,1].
\eeq

\end{itemize}

\subsubsection{Homology space}

\bp[Homology space \cite{Marcopath}]
The dimension of the $N=1$ homology space is
\beq
\dim H_1(e^{-V(x)}dx,K) = l+\deg V' = \sum_{p=\text{poles}} \deg_p V' + \# \text{hard edges}.
\eeq

\ep

\subsection{One matrix}

\subsubsection{Homology space}

With rational potentials, the homology space is not necessarily $\text{Sym}\left( H_1^{\otimes N}\right)$, it can be smaller.
$\EE:\Gamma \mapsto \EE_\Gamma$ is not necessarily injective.
We thus define
\beq
H_N = \text{Sym}\left( H_1^{\otimes N}\right)/\Ker\EE.
\eeq
$\Ker\EE\neq 0$ typically when $V'$ has simple poles with positive integer residues.
Let us write:
\bea
e^{-V(x)} =  \prod_{a=1}^s (x-p_a)^{-m_a} e^{-\td V(x)}
\eea
where $\td V'(x)$ has no simple pole with positive integer residues.

Let us illustrate what happens for $s=1$ and $N=2$ and $m_a=1$:
\bea
&& \int_{\mathcal C_{p_a}^2} (x-x')^2 p(x,x') \ \frac{1}{(x-p_a)(x'-p_a)} e^{-\td V(x)-\td V(x')} dx dx' \cr 
&=& 2\pi\ii e^{-\td V(p_a)}\  \int_{\mathcal C_{p_a}} (x-p_a)^2 p(x,p_a) \ \frac{1}{(x-p_a)} e^{-\td V(x)} dx  \cr 
&=& 0.
\eea
This means that $\mathcal C_{p_a}^2\in \Ker\EE$.

In general, computing integrals around $\mathcal C_{p_a}$ amounts to pinning one of the $x_i$s at $p_a$ and may diminish the degree of the pole at $p_a$ for the remaining $x_j$s, and if the degree drops too much, there is no poles and further integrations on $\mathcal C_{p_a}$ give zero.
In general we have:

\bp[Homology space (rational case)]
\beq
H_N = \text{Sym}\left( H_1^{\otimes N}\right)/\Ker\EE.
\eeq
\beq 
\label{eq:KerEcasrationel}
\Ker\EE = 
\sum_{\substack{1\le a\le s\\m_a<N}}
\mathcal C_{p_a}^{\,m_a+1}\,
\operatorname{Sym}^{N-m_a-1}H_1, 
\eeq
\beq
\label{eq:dimHNratl1MM}
\begin{aligned}
\dim H_N
&=\sum_{J\subseteq\{1,\ldots,s\}}(-1)^{|J|}
\binom{N+d+l-1-\sum_{a\in J}(m_a+1)}{d+l-1}\\
&=\frac{(N+d+l-1-\sum_{a=1}^s(m_a+1))!}{(d+l-1)!\,N!}
\sum_{J\subseteq\{1,\ldots,s\}}(-1)^{|J|}\\
&\qquad\times\prod_{i=1}^{\sum_{a\notin J}(m_a+1)}
\left(i+N+d+l-1-\sum_{a=1}^s(m_a+1)\right)\\
&\qquad\times\prod_{i=1}^{\sum_{a\in J}(m_a+1)}(N+1-i).
\end{aligned}
\eeq
(with the convention that $\binom{p}{q}=0$ if $p<q$, and rewriting in the second line of \eqref{eq:dimHNratl1MM} holds when $N+d+l-1-\sum_{a=1}^s(m_a+1)\geq 0$) .)

The map $\EE: H_N \to \mathcal P_N^*$ is an injective homeomorphism.

\ep

\proof{}
Let $P(x)=\prod_a(x-p_a)^{m_a}$.
Choose a subspace $H'_1\subset H_1$ mapping isomorphically onto
$H_1(e^{-\td V}dx)$. Thus
\[
H_1=H'_1\oplus\bigoplus_{a=1}^s\mathbb C\ \mathcal C_{p_a},
\qquad \dim H'_1=d+l-s.
\]

For $n$ variables integrated around $p_a$, residues extract the
coefficient of $\prod_{i=1}^n(x_i-p_a)^{m_a-1}$.
The Vandermonde squared in these variables has degree $n(n-1)$,
whereas the required coefficient has total degree $n(m_a-1)$.
Consequently all polynomial moments vanish if $n>m_a$.
This proves that the right hand side of \eqref{eq:KerEcasrationel} is included  in $\Ker\EE$.

Conversely, modulo this sum, write
\[
\Gamma=\sum_{\substack{0\le n_a\le m_a\\|\mathbf n|\le N}}
\mathcal C_{p_1}^{n_1}\cdots \mathcal C_{p_s}^{n_s}\,\Gamma_{\mathbf n},
\qquad
\Gamma_{\mathbf n}\in
\operatorname{Sym}^{N-|\mathbf n|}H'_1.
\]
Suppose $\Gamma\in\Ker\EE$ and choose a nonzero
$\Gamma_{\mathbf n}$ with $k=|\mathbf n|$ minimal.
Partition the first $k$ variables into groups $I_a$ of sizes $n_a$.
Choose polynomials $\ell_a$ such that $\ell_a(p_a)=1$ and
$\ell_a$ vanishes to order at least $m_b$ at $p_b$ for every $b\ne a$.
For any polynomial $q$ in the remaining variables, use the
symmetrization of
\beq
f_q(X)=
q(x_{k+1},\ldots,x_N)
\prod_{j=k+1}^N P(x_j)
\prod_a\prod_{i\in I_a}
(x_i-p_a)^{m_a-n_a}\ell_a(x_i).
\eeq
The factors $P(x_j)$ kill every circle factor assigned to a variable
$j>k$, while the $\ell_a$ select the prescribed poles in the first
$k$ variables. By minimality of $k$, only the chosen occupation
profile contributes.

Computing these residues gives
\beq
0=\EE_\Gamma(f_q)
=C_{\mathbf n}\,\EE^{(\mathbf n)}_{\Gamma_{\mathbf n}}(q),
\qquad C_{\mathbf n}\ne0,
\eeq
where the remaining weight is
\beq
e^{-V_{\mathbf n}(x)}
=e^{-\td V(x)}\prod_a(x-p_a)^{2n_a}.
\eeq
Indeed, for each group the relevant Vandermonde coefficient is
\beq
[u_1^{m_a-1}\cdots u_{n_a}^{m_a-1}]
\left(\Delta(u)^2\prod_i u_i^{m_a-n_a}\right)
=(-1)^{n_a(n_a-1)/2}n_a!\ne0.
\eeq
All other local factors contribute their nonzero constant terms.

The weight $e^{-V_{\mathbf n}}$ has no finite meromorphic poles.
Moreover, multiplying $e^{-\td V}$ by the displayed zeros only adds
admissible endpoints, so $H'_1$ embeds in its one-variable homology.

The rational injectivity lemma proved in Appendix~\ref{Appproofinjective},
Lemma~\ref{lemmarationalinjectivity}, applies to this remaining weight:
the added zeros give only negative integer residues of $V_{\mathbf n}'$,
and the regular hard edges remain admissible. It therefore implies
$\Gamma_{\mathbf n}=0$, a contradiction.
This proves the asserted description of $\Ker\EE$.

\smallskip

The quotient has a basis, made of circles $\mathcal C_{p_a}$ with occupation numbers 
$n_a\le m_a$,  and unrestricted occupations in the remaining $d+l-s$ basis of $H'_1$. 
With $\dim H_0=1$, the generating function computing the dimensions is
\beq
\begin{aligned}
\sum_{N\ge0}\dim H_N\,z^N
&=\frac{\prod_{a=1}^s\left(\sum_{n_a=0}^{m_a}z^{n_a}\right)}
        {(1-z)^{d+l-s}}\\
&=\frac{\prod_{a=1}^s(1-z^{m_a+1})}{(1-z)^{d+l}}.
\end{aligned}
\eeq
Extracting coefficients gives the stated dimension formula.
\eproof

For example is $s=1$ we get
\beq
\dim H_N = \binom{N+d+l-1}{d+l-1} - \binom{N+d+l-2-m_1}{d+l-1}.
\eeq

\subsubsection{Loop equations}

If there are hard-edges let
\beq
B(x) = \prod_{i=1}^l (x-b_i)
\eeq
and $B=1$ if there is no hard-edges.

Let us write $V'(x)$ as an irreducible rational fraction of 2 polynomials
\beq
V'(x) = \frac{R(x)}{D(x)}
\eeq
where $D(x)$ is a monic polynomial. Let us assume that $\deg R>\deg D$, and  we have
\beq
\deg R=d=\deg V'.
\eeq

Define the one-variable operator acting in $\CC[x]$
\beq
T p=(DB) p'+((DB)'-BR)p.
\eeq
Since the leading term of $T p$ has degree $d+l+\deg p$, every polynomial admits a unique decomposition
\beq
p=Tq+r,\qquad \deg r<d+l.
\eeq
Thus integration by parts gives a canonical linear reduction
\beq
p\mapsto \operatorname{red}(p)=r.
\eeq
For $N$ variables, consider the linear map on symmetric polynomials
\bea
L: & \mathcal P_N   \to & \mathcal P_N \cr 
& p(x_1,\dots,x_N)  \mapsto & \operatorname{red}_{x_1,\dots,x_N}(\Delta(X)^2 p).
\eea
We define
\beq
\mathcal L = \Ker L.
\eeq

\bd[Loop equations]
A linear form
$E\in \mathcal P_N^*$ is called a solution of loop equations iff
\beq
E(\mathcal L)=0.
\eeq
The set of solutions of loop equations is denoted
\beq
\mathcal L^\perp = \{E\in \mathcal P_N^* \ | \ \forall p\in \mathcal L, \ E(p)=0\}.
\eeq
\ed

\bt[Loop equations]
Every matrix integral is solution of loop equations:
\beq
\forall \Gamma \in H_N \ \qquad \EE_\Gamma(\mathcal L)=0
\eeq
i.e.
\beq
\operatorname{Im} \EE \subset \mathcal L^\perp .
\eeq

\et

\proof
For 1-variable we have
\beq
Tq(x) e^{-V(x)} = \frac{d}{dx}\left( B(x)D(x) q(x) e^{-V(x)} \right) 
\eeq
and since the integral of a total derivative of a polynomial proportional to $B(x)D(x)$ vanishes, the integral  on any $\gamma\in H_1$ of $p(x)$ and $r(x)=\operatorname{red}(p(x))$ are equal.
Therefore, if $p\in\Ker L$ we have
\beq
\int_\Gamma \Delta(X)^2 p(x_1,\dots,x_N) e^{-\Tr V(X)} dX
= \int_\Gamma L(p(x_1,\dots,x_N)) e^{-\Tr V(X)} dX = 0.
\eeq
i.e. $\EE_\Gamma(\Ker L)=0$.
\eproof

%
%

\bt
The map $\EE:H_N\to\mathcal L^\perp$ is an isomorphism. Consequently,
\beq
\dim\mathcal L^\perp=\dim H_N
=
\sum_{J\subseteq\{1,\ldots,s\}}
(-1)^{|J|}
\binom{N+d+l-1-\sum_{a\in J}(m_a+1)}{d+l-1}.
\eeq
\et

\proof
Let
\beq
\mathcal R= \mathbb C[x]_{<d+l}.
\eeq
For $N=1$, the one-variable period pairing
\beq
H_1\times\mathcal R \longrightarrow\mathbb C,
\qquad
(\gamma,r)\longmapsto
\int_\gamma r(x)e^{-V(x)}\,dx
\eeq
is perfect. 
Indeed, if a contour has zero pairing with every
$r\in \mathcal R$, reduction implies that all its polynomial
moments vanish. 
The one-variable injectivity result of
\cite{Bertola_2003,Marcopath}, for the admissible contours including
regular hard edges, then implies that the contour is zero.
Since $\dim H_1=\dim\mathcal R=d+l$, the pairing is perfect.

Taking symmetric tensor powers gives a perfect pairing between
\beq
K_N:=\operatorname{Sym}^N H_1
\qquad\text{and}\qquad
\mathcal R_N:=\operatorname{Sym}^N\mathcal R,
\eeq
given by
\beq
\langle\Gamma,r\rangle
=
\int_\Gamma r(X)e^{-\operatorname{Tr}V(X)}\,dX.
\eeq

Reduction commutes with permutations of the variables, hence
$L(\mathcal P_N)\subseteq\mathcal R_N$. Moreover, integration
by parts gives
\beq
\EE_\Gamma(p)=\langle\Gamma,L(p)\rangle,
\qquad \Gamma\in K_N,\quad p\in\mathcal P_N.
\eeq
Writing $W_N=L(\mathcal P_N)$, we therefore obtain
\beq
\ker\EE
=
W_N^\perp
:=
\{\Gamma\in K_N:
  \langle\Gamma,w\rangle=0\text{ for every }w\in W_N\}.
\eeq
By perfection of the pairing and finite-dimensional duality,
restriction to $W_N$ induces an isomorphism
\beq
H_N=K_N/\ker\EE
=K_N/W_N^\perp
\simeq W_N^*.
\eeq

On the other hand, $L$ induces an isomorphism
\beq
\mathcal P_N/\mathcal L\simeq W_N,
\eeq
so that
\beq
\mathcal L^\perp
\simeq(\mathcal P_N/\mathcal L)^*
\simeq W_N^*.
\eeq
Under these identifications the induced map is precisely $\EE$.
Thus $\EE:H_N\to\mathcal L^\perp$ is an isomorphism.
The dimension formula follows from the preceding proposition.
\eproof

%

\subsection{Chain of matrices}

The same proof generalizes immediately to chain of matrices with rational $V'_l\in \CC(x_l)$.

In the case where $e^{-V_j}$ have no poles (i.e. $V'_j$ has no simple poles with positive integer residues), we get that
\beq
\dim \mathcal L^\perp = \frac{(N+D-1)!}{N! (D-1)!}
\eeq
where $D=\prod_{j=1}^L (\deg V'_j+l_j)$ where $\deg V'_j$ is the sum of degrees of all the poles of $V'_j$ and $l_j$ the number of hard edges of the $j^{th}$ matrix.

\section{Examples of applications}

\subsection{Application: Haar measure on $U(N)$}

We already mentioned that if $\gamma=S^1$ the unit circle, we have
\beq
\mathcal H_N(S^1) = U(N)
\eeq
and the measure $\DD M$ is closely related to the Haar measure (see \cite{eynRMT}, it is easy to see that $\ii^{-N^2} \det M^{-N}\DD M$ is a real positive measure and is invariant under right or left multiplication by an element of $U(N)$ so is the Haar measure)
\beq
\ii^{N^2} \ \DD_{\text{Haar}}M = \frac{1}{(\det M)^{N}} \ \DD M = e^{-N\Tr \log M} \ \DD M ,
\eeq
so that the eigenvalues statistics of a random unitary matrix with Haar measure on $U(N)$, can be rewritten as a normal matrix whose potential is
\beq
V(x) = N \log(x)
\eeq
i.e. $V'$ a rational fraction
\beq
V'(x)=\frac{N}{x}
\quad , \quad
d=\deg V'=1.
\eeq
There is thus a unique homology class in $H_N$ which has dimension
\beq
\dim H_N = \frac{(N+d-1)!}{N!(d-1)!} = 1.
\eeq
This unique homology class is $(S^1)^N$, i.e. all eigenvalues are on the  circle.

\smallskip
We could also consider a Haar measure with polynomial potential of some degree $k+1$, typically
\beq
\left| e^{-\Tr V(M)} \right| \DD_{\text{Haar}}M
= e^{-\Tr (\frac12 V(M)+\frac12 V(M^{-1})+N\log M)} \ \DD M
\eeq
which is a normal matrix model with rational
\beq
\frac12 V'(x) - \frac{1}{2x^2}V'(1/x) + \frac{N}{x}
\eeq
of total degree
\beq
d=2k+2.
\eeq

\subsection{Application: normal matrices in the complex plane}

It is well known that one random complex matrix $M\in M_N(\CC)$, is closely related to one random normal complex matrix $M\in \mathcal H_N(\CC)$ and equivalent to a 2-matrix model. Let us recall how.

Consider a random complex matrix $M$ in $M_N(\CC)$, with a measure
\beq
e^{-\Tr M M^\dagger} e^{-\Tr (V(M)+\td V(M^\dagger))}  \DD_{\text{Lebesgue}} M
\eeq
where $\DD_{\text{Lebesgue}} M= \prod_{i,j} d\Re M_{i,j}\wedge d\Im M_{i,j} $ is the Lebesgue measure on $M_N(\CC)$.

For $V$ and $\td V$ polynomials of degree $\geq 3$ this integral may be non-convergent.
It can be absolutely convergent on $\CC$ for example for rational potentials like $V(x) = \sum_{j=1}^s (1-\alpha_j) \ln{(x-z_j)}$ and $\td V(x) = \sum_{j=1}^s (1-\alpha_j) \ln{(x-\bar z_j)}$, if $\alpha_j>0$.
However we shall see below that under a suitable replacement of $\CC$ by an appropriate homology domain, the integral can be convergent for any polynomials $V$ and $\td V$ of degree $\geq 3$.

Every complex matrix can be triangulized by a unitary transformation:
\beq
M = U(X+T)U^\dagger
\eeq
where
\beq
\begin{cases}
U \in U(N)/U(1)^N \cr 
X = \diag(x_1,\dots,x_N) \ , \quad x_i \in \CC \cr
T = \text{strictly upper triangular complex}
\end{cases}.
\eeq
The following Schur measure identities are understood up to constants
depending only on $N$. The measure is then
\beq
\DD_{\text{Lebesgue}} M = |\Delta(X)|^2 \DD X \ \DD T \ \DD U 
\eeq
where
\bea
\DD X &=& \prod_{i=1}^N d_{\text{Lebesgue}} x_i \cr 
\DD T &=& \prod_{i<j} d_{\text{Lebesgue}} T_{i,j} \cr 
\DD U &=& \text{Haar measure of }U(N)/U(1)^N.
\eea
Therefore
\bea
&& e^{-\Tr M M^\dagger} e^{-\Tr (V(M)+\td V(M^\dagger))}  \DD_{\text{Lebesgue}} M  \cr
&=& e^{-\Tr X X^\dagger - \Tr T T^\dagger} e^{-\Tr (V(X)+\td V(X^\dagger))}  \Delta(X) \Delta(X^\dagger) \DD X \ \DD T \ \DD U .
\eea

On the other hand, consider a normal random matrix $M$ in $\mathcal H_N(\CC)$, with a measure
\beq
e^{-\Tr M M^\dagger} e^{-\Tr (V(M)+\td V(M^\dagger))}  \DD M \DD M^\dagger
\eeq
where $ \DD M \DD M^\dagger$ denotes the measure on $\mathcal H_N(\CC)$ defined below. To define it, notice that both $M$ and $M^\dagger$ are normal matrices and can be diagonalized by the same unitary conjugation:
\beq
M=U X U^\dagger
\quad , \quad 
M^\dagger=U Y U^\dagger
\quad , \quad 
Y=\bar X
\eeq
where $U\in U(N)/U(1)^N$, and $X$ is a diagonal complex matrix.

The measure on $\mathcal H_N(\CC)$ is defined as
\beq
\DD M \DD M^\dagger
= |\Delta(X)|^2 \DD U \DD X \DD \bar X,
\eeq
where $\DD U$ is  the Haar measure on $U(N)/U(1)^N$ and
$\DD X \DD \bar X=\prod_{i=1}^N dx_i d\bar x_i$. Here
$dx_i d\bar x_i$ denotes the positive Lebesgue measure
$d\Re x_i\,d\Im x_i=\frac{\ii}{2}dx_i\wedge d\bar x_i$.

The integration over the triangular matrix $T$ factorizes out, with
\[
\int e^{-\Tr T T^\dagger}\,\DD T=\pi^{N(N-1)/2}.
\]
Thus the unnormalized polynomial moments
$\EE(\Tr M^{k_1}\dots \Tr M^{k_n})$ of a complex matrix in
$M_N(\CC)$ are proportional to those of a normal complex matrix in
$\mathcal H_N(\CC)$, with a factor depending only on $N$.
Their normalized moments coincide whenever normalization is defined.

\medskip

The induced marginal measure for eigenvalues is
\beq
 |\Delta(X)|^2   \prod_{i=1}^N e^{-(x_i \bar x_i + V(x_i)+\td V(\bar x_i))} dx_i d\bar x_i.
\eeq
It is a real measure when $\td V$ is the complex conjugate of $V$.

Considering $X$ and $Y=\bar X$ as independent variables we see that it is a 2-matrix model.
More precisely, it is a 2-matrix model where $X,Y$ are integrated on a
middle-dimensional real submanifold of $\CC^{2N}$, of real dimension
$2N$, satisfying $Y=\bar X$. If the integral is convergent, this manifold must be in $H_N$.
To pass from Lebesgue measure to the holomorphic form
$\prod_i(dx_i\wedge dy_i)$, we absorb the factor $(\ii/2)^N$ into
the integration class. Write $\Gamma$ for the resulting one-pair
class. The $N$-pair contour is its symmetrized $N$-fold product
\[
\Gamma^N\in H_N\left(\Delta(X)\Delta(Y)
\prod_{i=1}^N e^{-(x_i y_i+V(x_i)+\td V(y_i))}
dx_i\wedge dy_i\right).
\]
The coupling has the opposite sign to that in Section~3; the change
$y\mapsto-y$, with $\td V(y)$ replaced by $\td V(-y)$, gives the
convention used there.

For the remainder of this subsection, assume that \(V\) and \(\widetilde V\) are polynomials of degree at least \(3\).
If $\gamma_i$ (resp. $\td\gamma_i$) form a basis of
$H_1(e^{-V(x)}dx)$ (resp. $H_1(e^{-\td V(y)}dy)$), the one-pair
class $\Gamma\in H_1$ can be written as
\beq
\Gamma = \sum_{i=1}^{\deg V'}\sum_{j=1}^{\deg \td V'} c_{i,j} \gamma_i \times \td\gamma_j.
\eeq
If $\td V=\bar V$, we may choose $\td\gamma_i=\bar\gamma_i$.
Let $\sigma$ exchange $x$ and $y$. With the above Lebesgue phase
absorbed into the one-pair class, the reality condition is
$\bar\Gamma=-\sigma_*\Gamma$; for $N$ pairs it reads
$\overline{\Gamma^N}=(-1)^N\sigma_*(\Gamma^N)$, where $\sigma$
exchanges $X$ and $Y$.
Since oriented products satisfy
$\sigma_*(\gamma_i\times\bar\gamma_j)
=-\bar\gamma_j\times\gamma_i$, this condition is equivalent to
$c_{i,j}=\overline{c_{j,i}}$.
The matrix $c_{i,j}$ is thus Hermitian, and we can choose a basis in
which it is diagonal and real, i.e. we can choose
\beq
\Gamma = \sum_{i=1}^{\deg V'} c_i \gamma_i \times \bar\gamma_i \qquad , \ c_i\in \RR.
\eeq
Let us choose
\beq
\gamma_i = \zeta_i \RR_+ - \zeta_{i-1} \RR_+
\eeq
where $\zeta_j=t_{d+1}^{-\frac{1}{d+1}}
e^{2\pi\ii\frac{j}{d+1}}$ are roots of $t_{d+1}^{-1}$, and
$t_{d+1}$ is the leading coefficient of $V'$.
Thus $t_{d+1}\zeta_j^{d+1}=1$, so these rays lie in decay sectors
of $e^{-V(x)}$ at infinity.
$\gamma_1,\dots,\gamma_d$ form a basis of $H_1$ and we have
\beq
\gamma_{d+1}=-\sum_{i=1}^d \gamma_i.
\eeq
The following class
\beq
\Gamma = \sum_{i=1}^{d+1} \gamma_i\times \bar\gamma_i
\eeq
is (up to a real proportionality constant) a one-pair homology class
satisfying $\bar\Gamma=-\sigma_*\Gamma$ and invariant under
simultaneous rotations of $x$ by $2\pi/(d+1)$ and of $y$ by
$-2\pi/(d+1)$.
It is the natural candidate to replace $\CC$.

\subsection{Application: Combinatorics of maps}

See \cite{Berge2003,Eynbook, eynRMT} for an introduction to maps and random matrices
(Readers not familiar with combinatorics of maps may skip this part.)

Let $t_3,t_4,\dots t_{d+1}$ be complex numbers with $t_{d+1}\neq 0$, and $N\in \ZZ_+$. 
Let us denote the formal series $\hat T_{k_1,\dots,k_n}\in \mathbb Q[t_3,\dots,t_{d+1},N,N^{-1}][[t]]$

\beq
\begin{aligned}
\hat T_{k_1,\dots,k_n}
={}&\sum_{e=1}^\infty t^e
\sum_{m\in\mathbb M(e,k_1,\dots,k_n)}
\frac{N^{\chi(m)-n}}{\#\text{Aut}(m)}\\
&\qquad\times t_3^{n_3(m)}t_4^{n_4(m)}\dots t_{d+1}^{n_{d+1}(m)}.
\end{aligned}
\eeq
where $\mathbb M(e,k_1,\dots,k_n)$ is the (finite) set of connected orientable maps with $e$ edges, and made of $n_3$ triangles, $n_4$ quadrangles, $\dots$, $n_{d+1}$ $(d+1)$--angles, and with also $n$ marked labeled faces (a marked face is a face with a marked oriented edge on its boundary, so that the marked face is on the right of the marked edge) of respective size $k_1,k_2,\dots,k_n$.
We require $k_i\geq 1$, whereas unmarked faces have at least size 3 (triangles up to $(d+1)$--angles).
$\#\text{Aut}(m)$ is the automorphism factor of the map, $\#\text{Aut}(m)=1$ for maps with marked faces, and can be $\geq 1$ for $n=0$ (no marked faces).
$\chi(m) = \#\text{faces}(m)-\#\text{edges}(m)+\#\text{vertices}(m)$ is the Euler characteristic of the map.

For $e=1$, the only possibilities are one marked face of degree $2$
(a bridge) and two labelled marked faces of degree $1$ (a loop).
Their contributions are respectively $Nt$ to $\hat T_2$ and $t$
to $\hat T_{1,1}$; all other one-edge contributions vanish.

Let us define (again as formal power series of $t$)
\beq
T_{\emptyset}=e^{\hat T_{\emptyset}}
\eeq
and for $n\geq 1$
\beq
T_{k_1,\dots,k_n}
=e^{\hat T_{\emptyset}}
\sum_{\substack{\mu=\text{partition of}\\\{1,\ldots,n\}}}
\prod_{K\in\mu}\hat T_{(k_i)_{i\in K}}.
\eeq
The partitions are of the labels of the marked faces, so faces
of equal degree remain distinct.

For example:
\beq
T_{k_1} = T_{\emptyset} \hat T_{k_1} \ ,
\quad 
T_{k_1,k_2} = T_{\emptyset} (\hat T_{k_1,k_2}+\hat T_{k_1}\hat T_{k_2}), \ \dots
\eeq
It is well known that $T_{k_1,\dots,k_n}$ are generating functions for counting non--connected maps.

In the 1960's, W. Tutte \cite{Tutte1968, Tutte1} found some equations relating these generating functions, by recursion on the number of edges.
Tutte's equations can be rewritten as loop equations, let us explain how.

Let
\beq
V(x) = N\left( \frac{1}{2t}x^2 - \sum_{k=3}^{d+1} \frac{t_k}{k}x^k\right).
\eeq

Let $E:\mathcal P_N\to\CC[[t]]$ be the $\CC$-linear map defined
on the basis of power sum polynomials by
\beq
E(p_{\mu}) = T_{\mu_1,\dots,\mu_\ell}.
\eeq
When a zero index occurs, set
$T_{0,k_1,\ldots,k_n}=N T_{k_1,\ldots,k_n}$, since $p_0=N$.
We extend $E$ linearly over $\CC((t))$ when applying it to
polynomials with coefficients depending on $t$.

Wick's theorem gives the formal Gaussian representation
\eqref{eq:mapsformal} below \cite{Eynbook,eynRMT}.
Gaussian integration by parts applies coefficient by coefficient
to the expanded interaction. It gives Tutte's equations, which
are exactly the loop equations:
\beq
\forall \ \mu \ , \quad E(Q_\mu)=0 .
\eeq

More precisely, the generating functions have the representation
\beq\label{eq:mapsformal}
T_{k_1,\dots,k_n}
=\frac{\displaystyle\int_{\RR^N}^{\mathrm{formal}}
\left(\prod_{i=1}^n p_{k_i}(X)\right)
\Delta(X)^2 e^{-\Tr V(X)}\,dX}
{\displaystyle\int_{\RR^N}\Delta(X)^2
e^{-\frac{N}{2t}\Tr X^2}\,dX}.
\eeq
Here the numerator means that
$\exp\!\left(N\sum_{k=3}^{d+1}\frac{t_k}{k}p_k(X)\right)$
is expanded before Gaussian integration. The Gaussian integrals
are evaluated for $t>0$; their normalized ratios define formal
power series in $t$, with finitely many contributions at each degree.
The denominator is the Gaussian partition function, so
$T_\emptyset=1$ when all $t_k=0$.

Identity \eqref{eq:mapsformal} is an identity of formal series.
For a complex-valued realization of these moments that satisfies
the loop equations at fixed parameter values,
Theorem~\ref{mainth1} gives a contour representation over $\CC$, i.e. there exists some $\Gamma$ such that
\beq
T_{k_1,\dots,k_n}
= \EE_\Gamma\left(\prod_{i=1}^n \sum_{j=1}^N x_j^{k_i}\right).
\eeq

\section{Conclusion}

The theorems presented here are some "representation theorems", saying that linear forms on the space of symmetric polynomials, satisfying loop equations can always be represented as matrix-model-like measures (Vandermonde--square times exponential for the case of 1-matrix).
It also shows how normal matrices can be extremely useful.
We expect to prove similar theorems for the matrix model with external fields, or matrix models with hard edges.

Also we may guess some applications to free probabilities, to be explored further.

\section*{Acknowledgments and Declarations}

In its first version, this work was  supported by the ERC Synergie Grant ERC-2018-SyG  810573 "ReNewQUantum", and also partly supported by the ANR grant Quantact : ANR-16-CE40-0017.
I wish to thank IHES and M. Kontsevich, as well as University Paris Sud Orsay where I teach "random matrices" to Master students, and to whom I presented these loop equations theorem. I want to thank T. Kimura and S. Ribault who helped write this proof in my random matrix lecture notes \cite{eynRMT} given at IPHT in 2015.

\medskip

\begin{itemize}
\item \textbf{Funding:} The author is employed by IPHT/DRF/CEA. The version v1 was  supported by the ERC Synergie Grant ERC-2018-SyG  810573 "ReNewQUantum", and also partly supported by the ANR grant Quantact : ANR-16-CE40-0017. During the update to the second version v2, the author has not received external funding.

\item \textbf{AI use:} Several computations were tested with GPT-Astra 6.2 agents. The ideas are human, they are updated  from a version first posted in 2019.
AI was also used to polish the presentation.
\item \textbf {Competing interests:}
The author declares to have no competing interests.
\item \textbf {Data and code availability:}
No datasets were generated or analysed in this study.
\item \textbf {Ethics approval and consent: }Not applicable.

\end{itemize}

\appendix{}

\section*{Appendices}
\setcounter{section}{0}
\renewcommand{\thesection}{\Alph{section}}

\section{Lemma \ref{lemmaPN}}
\label{applemmaPN}

\textbf{Lemma \ref{lemmaPN}}:
{\em
A basis of $\mathcal P_N$ is given by 
\beq
\{ p_\mu \ | \ \ell(\mu)\leq N\}.
\eeq
Extension:
A basis of $\{ p \in \mathcal P_N \ | \ p \ \text{homogeneous of degree } d\}$ is given by 
\beq
\{ p_\mu \ | \ \ell(\mu)\leq N \ \text{and} \ |\mu|=d\}.
\eeq

}
\proof
By recursion on $N$. 
It is clearly true for $N=1$.

Assume it holds for $N-1$, let $P\in \mathcal P_N$ a symmetric polynomial of $N$ variables.
$P(x_1,x_2,\dots,x_N)$ can be expanded in powers of $x_1$
\beq
P(x_1,x_2,\dots,x_N) = \sum_k x_1^k Q_k(x_1,\dots,x_N)
\eeq
where each $Q_k \in \mathcal P_{N-1}$. 
By recursion  hypothesis there exists somecoeffisients $Q_{k,\nu}$
\beq
P(x_1,x_2,\dots,x_N) = \sum_k x_1^k \sum_{\nu,\ \ell(\nu)\leq N-1} Q_{k,\nu} p_{\nu_1}(x_2,\dots,x_N) \dots  p_{\nu_N}(x_2,\dots,x_N)
\eeq
Observe that
\beq
p_{\nu_i}(x_2,\dots,x_N) = p_{\nu_i}(x_1,x_2,\dots,x_N)-x_1^{\nu_i},
\eeq
therefore we can reexpand in powers of $x_1$
\beq
P(x_1,x_2,\dots,x_N) = \sum_k x_1^k \sum_{\nu,\ \ell(\nu)\leq N-1} \td Q_{k,\nu} p_{\nu_1}(x_1,\dots,x_N) \dots  p_{\nu_N}(x_1,\dots,x_N)
\eeq
By symmetry we also have $\forall \ i=1,\dots,N$
\beq
P(x_1,x_2,\dots,x_N) = \sum_k x_i^k \sum_{\nu,\ \ell(\nu)\leq N-1} \td Q_{k,\nu} p_{\nu_1}(x_1,\dots,x_N) \dots  p_{\nu_N}(x_1,\dots,x_N)
\eeq
and by summing over $i$
\bea
P(x_1,x_2,\dots,x_N) 
&=& \frac{1}{N}\sum_{i=1}^N \sum_k x_i^k \sum_{\nu,\ \ell(\nu)\leq N-1} \td Q_{k,\nu} p_{\nu_1}(x_1,\dots,x_N) \dots  p_{\nu_N}(x_1,\dots,x_N) \cr
&=& \frac{1}{N}\sum_k p_k(x_1,\dots,x_N) \sum_{\nu,\ \ell(\nu)\leq N-1} \td Q_{k,\nu} p_{\nu_1}(x_1,\dots,x_N) \dots  p_{\nu_N}(x_1,\dots,x_N) 
\eea
which is clearly a linear combination of $p_{\nu'}$ where $\nu'=\nu+(k)$ is a partition obtained by adding one part of length $k$ to $\nu$, and it has thus at most $N$ parts.
This proves the spanning assertion.

If $P$ is homogeneous of degree $d$, all steps preserve its degree,
so the homogeneous spanning assertion also holds.

Finally, in each degree $d$, the sums over distinct permutation
orbits of monomials form a basis: their monomial supports are
disjoint, and the coefficients of a symmetric polynomial are
constant on each orbit. These orbits are indexed by partitions
$\mu$ with $|\mu|=d$ and $\ell(\mu)\le N$. Thus the homogeneous
component has dimension equal to the number of $p_\mu$ in our
spanning family. This proves linear independence in each degree,
and hence both asserted basis statements.
\eproof

\section{Proof of injectivity theorem \ref{thinjective}}
\label{Appproofinjective}

\subsection{Polynomial case}
\label{AppproofinjectivePol}

Case where $V(x)\in K[x]$, and $H_N= \operatorname{Sym}^N H_1$:

\textbf{Theorem \ref{thinjective} (Polynomial injectivity)}

{\em
$\EE$ is an injective homeomorphism of vector spaces
\bea
\EE: & H_N &\to \mathcal P_N^* \cr
& \Gamma & \mapsto \EE_\Gamma
\eea 
 
}

\proof
Let us assume that $\Ker \EE\neq 0$, i.e. that there exists $\Gamma\neq 0$ such that $\EE_\Gamma(q)=0$ for all symmetric polynomial $q$.

Let us define the following linear form on $\CC[x_1,\dots,x_n]$:
\beq
L_\Gamma(q) = \int_\Gamma q(x_1,\dots,x_N) \prod_{i=1}^N e^{-V(x_i)} dx_i.
\eeq
Our assumption is that (by symmetrizing $q$):
\beq
\forall q\in \CC[x_1,\dots,x_n]
\qquad L_\Gamma(\Delta(X)^2 q)=0.
\eeq
Consider the ideal
\beq
I_\Gamma=
\{f\in \CC[x_1,\dots,x_n]:\ L_\Gamma(fq)=0\ \text{for every }q\in \CC[x_1,\dots,x_n]\}.
\eeq
It is not zero since it contains the Vandermonde square
\beq
\Delta^2 \in I_\Gamma.
\eeq
By integration by parts, if $f\in I_\gamma$ we have
\bea
L_\Gamma((\partial_i f ) q)
&=& L_\Gamma(f (V'(x_i) q - \partial_i q  ))
\eea
which implies that $\partial_i f\in I_\Gamma$.
By repeated differentation of the Vandermonde squared, we obtain that
\beq
1\in I_\gamma
\eeq
and thus
\beq
I_\Gamma = \CC[x_1,\dots,x_n].
\eeq
This implies that $L_\Gamma=0$.

Now we use the 1-variable injectivity:
\bl For $N=1$, the map $\mathbb E: \gamma \mapsto E_\gamma$ is injective.
\el
proved for instance in \cite{Bertola_2003}.

This implies that the vectors 
\beq
\left(\int_{\gamma_a}p(x)e^{-V(x)}dx\right)_{a=1}^d,
\qquad p\in\mathbb C[x],
\eeq
span $\CC^d$.
We can thus choose polynomials \(q_1,\ldots,q_d\) such that
\beq
\int_{\gamma_a}q_b(x)e^{-V(x)}dx=\delta_{ab}.
\eeq
And thus, if $\Gamma = \sum_{i_1,\dots,i_N} c_{i_1,\dots,i_N} \gamma_{i_1}\times \dots \times \gamma_{i_N}$, we have
\bea
c_{i_1,\dots,i_N} = \int_\Gamma q_{i_1}(x_1)\dots q_{i_n}(x_n) e^{-V(x_1)}dx_1 \dots e^{-V(x_N)}dx_N=  L_\Gamma(q_{i_1}\dots q_{i_N}) =0
\eea
which vanishes because $L_\Gamma=0$.
This implies that $\Gamma=0$, thus contradicting our non-injectivity hypothesis.
This proves that $\mathbb E$ is injective.
\eproof

\subsection{Rational case}
\label{AppproofinjectiveRat}

We now consider the same adapted lemma when $V'(x)\in K(x)$ is rational.

\bl[Rational injectivity]
\label{lemmarationalinjectivity}
For rational $V'$ and the admissible contours of
Section~\ref{secrational}, suppose that no finite simple pole of $V'$
has a positive integer residue. Regular hard edges are allowed.
Then the map
\beq
\EE:\operatorname{Sym}^N H_1\longrightarrow\mathcal P_N^*,
\qquad \Gamma\longmapsto\EE_\Gamma,
\eeq
is injective.
\el

\proof
Use $D,B,T$ from Section~\ref{secrational} and $L_\Gamma,I_\Gamma$
from the  proof in section \ref{AppproofinjectivePol} above.
Assume that there exists $\Gamma\in  \operatorname{Sym}^N H_1$ such that $\EE_\Gamma=0$. 
As above, symmetrizing a test polynomial gives $\Delta^2\in I_\Gamma$.

For $f\in I_\Gamma$, integration by parts gives
\beq
L_\Gamma\bigl((DB)(x_i)(\partial_i f)q\bigr)
=-L_\Gamma(f\,Tq)=0,
\eeq
where $T$ acts in the variable $x_i$ and $q$ is any polynomial.
The boundary terms vanish because $DB\,e^{-V}$ vanishes at every
admissible open end; the factor $B$ ensures vanishing of the boundary terms at  hard edges. 
Since $Tq$ is polynomial, we have proved
\beq
(DB)(x_i)\partial_i I_\Gamma\subseteq I_\Gamma.
\eeq

Let us prove that this implies
\bl
There exists some integer $M\geq 1$ such that
\beq
\prod_{i=1}^N\bigl(D(x_i)B(x_i)\bigr)^M\in I_\Gamma.
\eeq
\el

\proof{}
Consider the localized ring
\beq
R=\CC\left[x_1,\ldots,x_N,
\frac{1}{\prod_j D(x_j)B(x_j)}\right]
\eeq
and let $\td I_\Gamma$ be the ideal generated by $I_\Gamma$ in $R$.
The elements of $\td I_\Gamma$ are fractions $f/q$,
where $f\in I_\Gamma$ and $q$ is a product of nonnegative powers of
$D(x_j)B(x_j)$. 
Each individual factor becomes invertible, since
\beq
\frac{1}{D(x_i)B(x_i)}
=\frac{\prod_{j\ne i}D(x_j)B(x_j)}
       {\prod_j D(x_j)B(x_j)}.
\eeq
For any such fraction
\beq
\partial_i\left(\frac{f}{q}\right)
=\frac{(DB)(x_i)\partial_i f}{(DB)(x_i)q}
 -\frac{f\,\partial_i q}{q^2}.
\eeq
The numerator of the first fraction belongs to $I_\Gamma$ by the
inclusion just proved. The numerator of the second belongs to
$I_\Gamma$ because it is an ideal and $\partial_i q$ is polynomial.
Both denominators are allowed in the ring above. 

This implies that
\beq
\partial_i \td I_\Gamma \subset \td I_\Gamma.
\eeq
Since $\Delta^2\in \td I_\Gamma$, 
repeated differentiation of $\Delta^2$ implies that $1\in \td I_\Gamma$, which is equivalent to say that there exists some integer $M\ge1$ such that
\beq
\prod_{i=1}^N\bigl(D(x_i)B(x_i)\bigr)^M\in I_\Gamma.
\eeq
\eproof

\medskip

We now show that $T$ is invertible on the finite-dimensional space
$\CC[x]/(DB)^M\CC[x]$. First, $T$ preserves the ideal
$(DB)^M\CC[x]$, so it induces an operator on this quotient.
By the Chinese remainder theorem, it suffices to check the factors
at the zeros $p$ of $DB$. In the local basis
$1,(x-p),(x-p)^2,\ldots$, its matrix is triangular, with diagonal
entries
\beq
\begin{cases}
(DB)'(p)\bigl(j+1-\Res_p V'\bigr),
&\text{if $p$ is a simple zero of $DB$},\\
-B(p)R(p),
&\text{if $p$ is a multiple zero of $DB$}.
\end{cases}
\eeq
Here $j$ is a nonnegative integer in the relevant finite local basis.
At a hard edge $\Res_p V'=0$. At a simple pole, the first entry
cannot vanish by the hypothesis on residues. At a multiple zero,
$p$ is a multiple pole of $V'$, and $B(p)R(p)\ne0$ because
$R/D$ is reduced and the hard edges are distinct from the poles.
Thus all diagonal entries are nonzero, proving invertibility.

Consequently, every one-variable polynomial admits a decomposition
\beq
p=Tq+(DB)^M r.
\eeq
Integration by parts removes the $Tq$ term. Applying this
decomposition successively in each variable, for any polynomial
$p(x_1,\ldots,x_N)$, modulo integration by parts we obtain a polynomial $r(x_1,\ldots,x_N)$ with
\beq
L_\Gamma(p)
=L_\Gamma\left(\prod_{i=1}^N
       \bigl(D(x_i)B(x_i)\bigr)^M r\right)=0
\eeq
which vanishes because by our assumption $L_\Gamma$ vanishes on $I_\Gamma$.
Therefore $L_\Gamma=0$.

The end of the proof is the same as the end of the proof in the polynomial case in section \ref{AppproofinjectivePol}:
One-variable injectivity establishes an exact pairing and gives a basis of 1-variable polynomials dual to any basis of $H_1$. 
Their products recover all coefficients of $\Gamma$, exactly as at the end of the preceding proof. These coefficients must vanish because $L_\Gamma=0$.
Hence $\Gamma=0$, which concludes injectivity.
\eproof

\end{document}